\documentclass[twocolumn,amsmath,amssymb]{revtex4}

\usepackage{graphicx}
\usepackage{dcolumn}
\usepackage{bm}

\begin{document}


\title{Two-dimensional Hawking radiation from the AdS/CFT correspondence}

\author{Jorge S. D\'iaz}
\email{jsdiazpo@indiana.edu} %
\affiliation{Physics Department, Indiana University, Bloomington, Indiana 47405, USA }%


\begin{abstract}
The AdS/CFT correspondence has been tested through the
reproduction of standard results. Following this approach, we use
the correspondence to obtain the Hawking temperature of a black
hole in 1+1 dimensions. Using an auxiliary Liouville field, the
holographic energy-momentum tensor is found and compared with a
radiation energy-momentum tensor, verifying that the
correspondence gives the correct temperature. The information
about the radiated field in the CFT sector is contained in the
central charge, whereas in the radiation tensor this information
is in the statistical distribution. This result allows to
determine the radiation of scalar and Dirac fields easily and
without the necessity of solving the corresponding Klein-Gordon or
Dirac equation in a curved spacetime. In both cases, the correct
temperature was obtained.
\end{abstract}

\maketitle

\section{Introduction}

Since the establishment of the two governing theories of modern
physics, Quantum Mechanics and the General Theory of Relativity, a
search has been underway for a unification of these theories. The
goal of which would be understanding the short-range behavior of
gravity. The result, a quantum theory of gravity, would be key to
understanding such phenomena as the behavior of particles during
the early universe, or the mechanism of Hawking radiation
\cite{Hawking}.

Over the last few decades, superstring theory has became the most
promising framework to describe a unified picture of the known
interactions. One of the breakthroughs of last ten years in this
field was the AdS/CFT correspondence \cite{Maldacena}. This
duality establishes a one-to-one correspondence between the states
of a five-dimensional superstring theory in a Anti-de Sitter
(AdS$_5$) spacetime and a four-dimensional conformal field theory
(CFT$_4$) laying at the boundary of the AdS manifold. This can be
extended to a different dimensions as an AdS$_{d+1}$/CFT$_d$
correspondence. In the low energy limit the superstring theory
reduces to a gravitational theory, thus AdS/CFT becomes a
dictionary between a ($d+1$)-dimensional gravitational theory and
a $d$-dimensional quantum theory. In this sense, AdS/CFT offers
the possibility to understand quantum aspects of gravity. This
duality has been verified only on a case-by-case basis and until
now, though mathematically quite appealing, remains as a
conjecture. Several authors have used the correspondence to
reproduce standard results in order to try to understand its
origins as well as test the consistency of the conjecture. In this
vein, we set out to determine the Hawking temperature of a
two-dimensional black hole is determined through the holographic
energy-momentum tensor calculated using AdS$_3$/CFT$_2$.

In the next section, the procedure to identify the temperature
from the energy-momentum tensor and how the AdS/CFT correspondence
is used to determine this tensor is discussed. Section III
contains the explicit calculation of the holographic tensor;
whereas the calculation of the radiation tensor is presented in
section IV. In the final section the results are summarized and a
possible extension to a four-dimensional black hole is outlined.

\section{The energy-momentum tensor}

\subsection{The AdS/CFT correspondence and the energy-momentum tensor}

In quantum field theory, the partition function is written as a
path-integral over all fields $\Phi$ as follows

\begin{equation} \mathcal{Z}=\int\!\mathcal{D}\Phi\,e^{I[\Phi]}, \end{equation}
which is a useful function to determine expectation values of
physical quantities by coupling the desired quantity to its
corresponding current in the action $I[\Phi]$. The object of our
interest is the energy-momentum tensor, which can be obtained by
coupling the action to the metric

\begin{equation}\label{<T>} \langle\, T_{ij}\rangle =\frac{\delta\mathcal{Z}}{\delta g^{ij}}
=\int\!\mathcal{D}\Phi\,T_{ij}\,e^{I[\Phi,g]}.  \end{equation} %

One aspect of the AdS/CFT correspondence establishes that the
exponential of the $(d+1)$-dimensional gravity action as a
function of the induced metric of its boundary is exactly the
partition function of a $d$-dimensional conformal field theory
lying on that boundary spacetime

\begin{equation}\label{AdS_CFT}
e^{I_{\!\mbox{\tiny AdS}}[g_{(0)}]}=\int\!\mathcal{D}\Phi\,e^{I_{\!\mbox{\tiny CFT}}[\Phi,g_{(0)}]}.%
\end{equation}

In order to understand the way to use this correspondence, lets
look at the left side in more detail. In this work we shall study
a two-dimensional quantum theory and its correspondence to a
three-dimensional dual gravity theory. In the low energy limit,
the superstring theory in AdS$_3$ reduces to Einstein's equations
with negative cosmological constant, hence the three-dimensional
metric satisfies

\begin{equation}\label{Einstein eq} G_{\mu\nu}+\Lambda\,g_{\mu\nu}=0,\end{equation}
where the cosmological constant is related with AdS$_3$ radius by
$\Lambda=-l^{-2}$, and Greek indices are used for the bulk (2+1
dimensional) solution, whereas Latin indices are used for boundary
induced (1+1 dimensional) spacetime. The solution of the
differential equation (\ref{Einstein eq}) is the bulk-metric
$g_{\mu\nu}(z,x)$, which can be written as a function of its
boundary condition determined by the two-dimensional metric
$g_{(0)ij}(x)$. Therefore, the three-dimensional action is reduced
to a two-dimensional action

\begin{equation} I_3[g(g_{(0)})]\rightarrow I_2[g_{(0)}], \end{equation}
thus, using (\ref{<T>}) and (\ref{AdS_CFT}), the expectation value
of the energy-momentum tensor describing the quantum theory can by
obtained varying the gravitational action with respect to the
boundary metric

\begin{equation} \langle\, T_{ij}\rangle _{\!\mbox{\tiny CFT}}=\frac{\delta I_2}{\delta g^{ij}_{(0)}}. \end{equation} %

\subsection{1+1 Black hole temperature and the energy-momentum tensor}

In two dimensions, the conformal anomaly and the energy-momentum
conservation

\begin{equation}\label{conformal_anomaly}
T^\mu_{\phantom{\mu}\mu}=\frac{c}{24\pi}\,R\qquad,\qquad \nabla_\mu T^\mu_{\phantom{\mu}\nu}=0
\end{equation} %
completely determine the regularized energy-momentum tensor. By
comparing this tensor with that representing a radiation flux, the
temperature can be identified. This procedure has been followed by
several authors \cite{Christensen-Fulling,Ale,Wipf} for massless
scalar fields and the temperature found agrees with the one
predicted by Hawking \cite{Hawking}.

In this work the procedure for finding the temperature by
comparing the energy-momentum tensor of a quantum field in a black
hole background with the one representing a thermal flux is the
same as followed by authors above; nonetheless, the way to
calculate the expectation value of the energy-momentum tensor will
be by invoking AdS/CFT.

\section{Holographic energy-momentum tensor}

The three-dimensional AdS spacetime will be described by a element
of line written in Fefferman-Graham coordinates \cite{FG}, as
follows

\begin{equation}\label{Coord_FG1} ds^2=\frac{l^2}{z^2}\biggr(
dz^2+g_{ij}(x,z)\,dx^idx^j\biggr). \end{equation}

These coordinates allow for the expansion of the two-dimensional
metric $g_{ij}(x,z)$ as power series

\begin{equation}\label{g_serie(d)} g_{ij}(x,z)=g_{(0)ij}(x)+z^{2}\,g_{(1)ij}(x)+z^4\,g_{(2)ij}(x).\end{equation} %
Generically this expansion is infinite; nevertheless, in three
dimensions the Weyl tensor always vanishes hence the metric is
conformally flat and the FG expansion becomes finite.
Additionally, as the boundary of the three-dimensional manifold is
defined at $z=0$, this expansion allows for the identification the
induced two-dimensional metric $g_{(0)ij}$.

It was shown in \cite{Skenderis(Holographic Reconstruction)} that
the AdS/CFT correspondence gives, after the regularization of the
gravitational action, the expectation value of the energy-momentum
tensor in terms of the coefficients of the FG expansion as

\begin{equation}\label{T_ij} \langle\, T_{ij}\rangle =\frac{l}{8\pi G_3}\biggr(
\,g_{(1)ij}-g_{(0)ij}\,\mbox{Tr}\,g_{(1)}\biggr), \end{equation}
where $G_3$ is the Newton constant in three dimensions. The
energy-momentum tensor of the conformal theory is completely
determined by the metric induced at boundary $g_{(0)}$. In
general, all terms $g_{(k)}$ can be written in terms of $g_{(0)}$
because this first coefficient of the FG expansion is the boundary
condition for equation (\ref{Einstein eq}). For the particular
case of $k=d/2$ this dependence is non-local and when $d=2$
Einstein's equations only fix its trace \cite{Skenderis(Quantum
Effective)}:

\begin{equation}\label{Tr(g_1)[1+1]} \mbox{Tr}\,g_{(1)}=\frac{1}{2}\, R_{(0)}, \end{equation}
where $R_{(0)}$ is the curvature of $g_{(0)}$. This property was
used by Skenderis and Solodukhin \cite{Skenderis(Quantum
Effective)} to introduce a Liouville field as an auxiliary field,
in which case the energy-momentum tensor reads

\begin{multline}\label{T_ij(Liuoville)}
\langle\, T_{ij}\rangle=\frac{l}{16\pi G_3}\left[\frac{1}{2}\,\nabla\!_i\phi\,\nabla\!_j\phi-\nabla\!_i\nabla\!_j\phi\right. %
\\ \left.+\,g_{(0)ij}\biggr(\square\phi-\frac{1}{4}\,\nabla^k\phi\nabla_k\phi\biggr)\right],\end{multline}
where the covariant derivative is taken using $g_{(0)}$ and the
auxiliary field $\phi$ satisfies the Liouville field equation
without potential

\begin{equation}\label{Liouville_field_eq} \square\phi=R_{(0)}. \end{equation}

The three-dimensional Newton constant $G_3$ and the AdS radius $l$
can be related to the central charge of the CFT through the
Brown-Henneaux \cite{Brown&Henneaux} central charge $c=3l/2G_3$.
Thus the energy-momentum tensor only depends on the boundary
metric and the CFT is characterized by the central charge. With
this the correct conformal anomaly (\ref{conformal_anomaly}) can
be obtained \cite{HenningsonSkenderis(1), HenningsonSkenderis(2)}.
This energy-momentum tensor then becomes the same that is found in
\cite{DFU} for a scalar field when $c=1$ is chosen.

In order to describe the energy-momentum tensor in a black hole
background, the boundary metric is written as

\begin{equation}\label{ds2_2D} ds^2_{1+1}=g_{(0)ij}\,dx^idx^j=-f(r)\,dt^2+\frac{dr^2}{f(r)}. \end{equation}%

This metric features an event horizon $r=r_+$ with $f(r_+)=0$ and
$f'(r_+)\neq0$. The two integration constants obtained when
(\ref{Liouville_field_eq}) is solved are fixed by imposing
regularity of the energy-momentum tensor in the future horizon
$\mathcal{H}^+$, in order to get a particle flow at infinity which
is only described by the Unruh vacuum state \cite{Ale}. After
fixing these constants it is possible to show that, if the metric
(\ref{ds2_2D}) is asymptotically flat at infinity, all the
components of the tensor (\ref{T_ij(Liuoville)}) converge to the
same value given by

\begin{equation}\label{T_ij (infty)_(c)}
\langle\, T^{ij}\rangle =\frac{c\,f'^2(r_+)}{192\pi}, \qquad i,j=0,1
\end{equation} %
which is the tensor needed to identify the temperature of a flux
of particles at infinity. In the works mentioned above
\cite{Christensen-Fulling,Ale,DFU} the energy-momentum tensor was
calculated for a massless scalar field, whose central charge is
$c=1$. An important feature of result (\ref{T_ij (infty)_(c)}) is
the presence of the central charge, because it distinguishes
between scalar and Dirac fields when it is treated as the
parameter of the theory \cite{BPZ}.


\section{Radiation energy-momentum tensor and Hawking temperature}

In this section we show the calculation of the energy-momentum
tensor of a radiated field with statistical distribution
$\langle\, n_\omega\rangle $. It is possible to show that for
massless fields the magnitude of energy density, flux, and
radiation pressure are exactly the same ($\rho=F=P$); therefore,
all the energy-momentum tensor components are equal $T^{ij}=\rho$
\cite{yo}. In two dimensions, the energy density for a given type
of field is

\begin{equation}
\rho=\frac{1}{2\pi}\int_0^\infty \omega\,\langle\, n_\omega\rangle \,d\omega.
\end{equation} %

If we have $\Upsilon$ fields, the components of the total
radiation energy-momentum tensor will be simply given by

\begin{equation}\label{T_rad}
T^{ij}=\frac{\Upsilon}{2\pi}\int_0^\infty \omega\,\langle\, n_\omega\rangle \,d\omega, \qquad i,j=0,1
\end{equation}%

Below, this result will be used to calculate the temperature of
the radiation of massless scalar and Dirac fields.

\subsection{Thermal radiation of massless scalar fields}

In order to identify the temperature of the two-dimensional black
hole radiating scalar fields, we compare (\ref{T_ij (infty)_(c)})
and (\ref{T_rad}):

\begin{equation}\label{T_eq(scalar)}
\frac{c\,f'^2(r_+)}{192\pi}=\frac{\Upsilon}{2\pi}\int_0^\infty
\omega\,\langle\,n_\omega^{\mbox{\tiny BE}}(T)\rangle \,d\omega,
\end{equation} %
where the temperature $T$ is contained in the Bose-Einstein
distribution $\langle\, n_\nu^{BE}(T)\rangle $. This last equation
shows that presence of the central charge in (\ref{T_ij
(infty)_(c)}) is essential because it is somehow a measure of the
number of degrees of freedom of the system \cite{CFT}. As it is
additive, the central charge of a system of scalar fields is
exactly the number of them $c=\Upsilon$, and hence the temperature
is

\begin{equation}\label{T_H} T=\frac{1}{4\pi}\,f'(r_+).\end{equation}
%
%

\subsection{Thermal radiation of massless Dirac fields}

In his original paper, Hawking \cite{Hawking} claims that besides
scalar fields, a black hole could also radiate massless fermionic
fields. In spite of the fact that fermions obey a different
distribution, the same temperature (\ref{T_H}) would be found.
Several works have explicitly shown this result from different
approaches: the uniformly accelerated detector in vacuum method,
shown separately by Davies \cite{Davies_accelerated} and Unruh
\cite{Unruh_accelerated}, was extended to treat the case where the
field seen by the accelerated observer is a spin-1/2 Dirac field
\cite{accelerated_Alsing}; the Dirac equation was studied in a
black hole background and provides a derivation of the Hawking
temperature \cite{Dirac_eq_curved_st}; and in the
three-dimensional black hole the Hawking radiation of Dirac fields
agrees with the one obtained from the scalar field case
\cite{Dirac_fields_2+1}. In our approach, the information about
the type of fields used for AdS/CFT calculations is in the central
charge. The central charge of one Dirac field is 1/2, hence each
field contribute with this quantity to the total central charge;
so for $\Upsilon$ Dirac fields we have $c=\Upsilon/2$. Using this
and replacing Bose-Einstein by Fermi-Dirac distribution, equation
(\ref{T_eq(scalar)}) becomes

\begin{equation}\label{T_eq(Dirac)}
\frac{f'^2(r_+)}{384\pi}=\frac{1}{2\pi}\int_0^\infty
\omega\,\langle\,n_\omega^{\mbox{\tiny FD}}(T)\rangle\,d\omega.
\end{equation} %

Even though the left hand side decreases by a half compared with
the scalar case (\ref{T_eq(scalar)}), the temperature is the same
because in (\ref{T_eq(Dirac)}) we use the Fermi-Dirac distribution
and the integral on the right hand decreases by a half as well.
Therefore, the temperature from (\ref{T_eq(Dirac)}) is also given
by (\ref{T_H}).

\section{Conclusions}

AdS/CFT correspondence allows computation of the Hawking
temperature for a 2D black hole by direct comparison of both sides
of the duality. In the AdS gravity side,  the finite boundary
energy-momentum tensor was obtained using the Fefferman-Graham
expansion. The induced theory on the boundary is an auxiliary
Liouville field whose stress tensor depends of the metric and the
central charge of the two-dimensional CFT. As this number
characterizes the field theory it contains the information about
the number and the type of fields radiated by the black hole. The
thermal energy-momentum tensor was found, which has the
information about the fields radiated in the statistical
distribution. By comparing the energy-momentum tensor calculated
by both methods, it was possible to identify of the black hole
temperature. The method presented in this paper allows for the
determination of the temperature of a black hole radiating
massless scalar fields and Dirac fields without the necessity of
solving the corresponding Klein-Gordon and Dirac equations on a
curved spacetime. In both cases the temperature agrees with the
one found by Hawking.

The next step is the extension of the procedure shown in this work
to a four-dimensional black hole. An important difference will
appear for the AdS$_5$/CFT$_4$ calculations because, for a given
string theory, the correspondence establishes exactly what the CFT
is. For instance, Type IIB string theory on AdS$_5$ has a
corresponding CFT which is the $\mathcal{N}=4$ super-Yang-Mills
gauge theory. The number of fields is also already determined:
there are six scalar fields, two Dirac fields, and one Yang-Mills
field; in the large $N$ limit, each of the $N^2$ parameters of
SU($N$) contribute with a degree of freedom; therefore, each field
will contribute with a factor $N^2$ to the radiation
energy-momentum tensor. Additionally, the relation between the AdS
radius and the central charge used in the two-dimensional case
would be replaced by parameters as fundamental as the length of
the string; therefore, a non-trivial test of AdS/CFT would be its
capability to provide the temperature for a four-dimensional black
hole.

\begin{acknowledgments}

I would like to thank M. Ba\~nados for his advice and many
enlightening conversations. Also thanks to A. Castro, whose work
was crucial to understand subtle details of this project. Special
thanks to R. Olea, A. Reisenegger, and J.P. Staforelli, for
discussions which contributed enormously to this work. Important
remarks on an earlier version of the manuscript by M. Ahmad, M.
Berger, M.J. Cordero, V.A. Kosteleck\'y, and N. Poplawski, are
also gratefully acknowledge. Finally, thanks to Centro de Estudios
Cient\'ificos in Valdivia, Chile for hospitality during the
initial stages of this work.

\end{acknowledgments}

\newpage 

\end{document}